\documentclass{Interspeech}

\interspeechcameraready 
\usepackage{cite}
\usepackage{amsmath,amssymb,amsfonts,booktabs,multirow}
\usepackage{multicol}
\usepackage{algorithmic}
\usepackage{graphicx}
\usepackage{textcomp}
\usepackage{xcolor}
\usepackage{arydshln}
\usepackage{lineno,hyperref}
\usepackage[obeyspaces]{xurl}

\title{MDDM: A Multi-view Discriminative Enhanced Diffusion-based Model for Speech Enhancement}

\author{Nan}{Xu}
\author{Zhaolong}{Huang}
\author{Xiaonan}{Zhi}

\affiliation[nocounter]{Alibaba Digital Media Entertainment Group}{Beijing}{China}
\email{\{xn430658,zhaolong.hzl,dexiao.zxn\}@alibaba-inc.com}

\usepackage{comment}

\begin{document}

\maketitle

\begin{abstract}
 \sloppy
With the development of deep learning, speech enhancement has been greatly optimized in terms of speech quality. Previous methods typically focus on the discriminative supervised learning or generative modeling, which tends to introduce speech distortions or high computational cost. In this paper, we propose {\bf MDDM}, a {\bf M}ulti-view {\bf D}iscriminative enhanced {\bf D}iffusion-based {\bf M}odel. Specifically, we take the features of three domains (time, frequency and noise) as inputs of a discriminative prediction network, generating the preliminary spectrogram. Then, the discriminative output can be converted to clean speech by several inference sampling steps. Due to the intersection of the distributions between discriminative output and clean target, the smaller sampling steps can achieve the competitive performance compared to other diffusion-based methods. Experiments conducted on a public dataset and a real-world dataset validate the effectiveness of MDDM, either on subjective or objective metric.
\end{abstract}

\section{Introduction}
\begin{figure*}[t]
\centering
\includegraphics[width=0.94\textwidth]{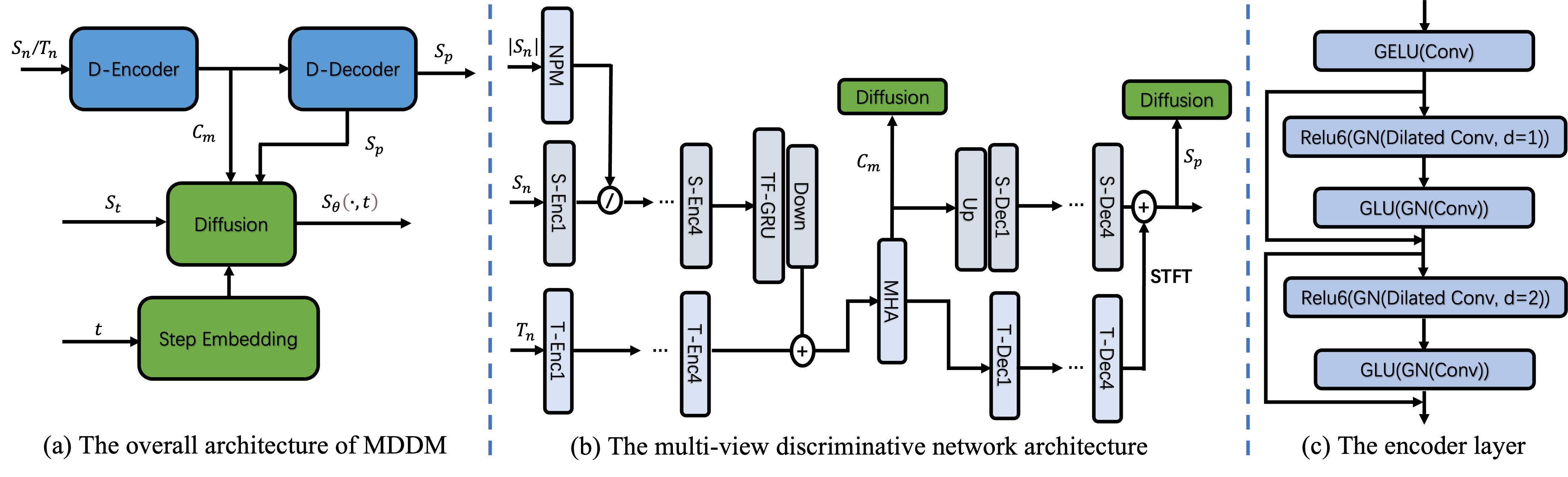}
\caption{The overall architecture of MDDM, multi-view discriminative network and encoder layer. In subfigure (a), $ S_n $ and $ T_n $ are noisy signals in frequency and time domains, respectively. Two D-* are encoder and decoder of discriminative network and output spectrogram is $ S_p $.  $ S_t $ is the sampled spectrogram at time-step $ t $. The intermediate multi-view feature $ C_m $ and output $ S_p $ are the conditions of diffusion model. In subfigure (b), S-* and T-* means frequency and time domains, respectively. NPM is the noise perception module and $ \oslash $ means the modulation mode. Note that the skip connections are not drawn. In subfigure (c), the encoder module is shown. GELU is Gaussian Error Linear Unit and GN denotes GroupNorm. The dilation rates are 1 and 2, respectively.}
\label{frmw}
\end{figure*}

\sloppy
In the real world, clean speech signals are always disturbed by various environmental noises and reverberations, which seriously affects speech perceptual quality and intelligibility. Therefore, speech enhancement is thus an underlying task, which aims to recover clean speech signal from noisy speech. Traditional speech enhancement methods can use the statistical properties of the noisy and clean target signals in time-frequency or spatial domain \cite{gerkmann2018spectral}. With the success of deep learning, the enhancement performance has achieved some breakthroughs in the last decade. Deep learning based methods can be divided into two different categories: discriminative approaches and generative approaches.

\sloppy
Discriminative approaches are dominated by supervised learning algorithms that obtain clean target speech from noisy speech, always training with labeled samples. These methods typically take time domain, time-frequency (TF) domain or both as network input to learn a deterministic mapping. Specifically, the time or time-frequency domain methods take the single-view feature as input to predict waveform or magnitude-phase related training target \cite{stoller2018wave,luo2019conv,hu2020dccrn,li2020noise,yu2022high}. Instead, some approaches integrate time domain and time-frequency domain in the speech enhancement framework, primely eliminating non-stationary (e.g., impulse-like noise) and stationary noises simultaneously \cite{zhang2021dbnet,defossez2021hybrid,rouard2023hybrid}. In addition, several methods follow the ideas of generative models for speech enhancement, such as variational autoencoder (VAE) \cite{leglaive2018variance,richter2020speech,bie2022unsupervised} or Generative adversarial networks (GANs) \cite{routray2022phase,liu2021voicefixer}. In contrast, diffusion-based speech enhancement models have recently gained attention due to the superior enhancement effects \cite{tai2023revisiting,guo2023variance,hu2024noise,tai2024dose}. Diffusion-based enhancement framework usually adds noise from the Wiener process to make clean target speech into a tractable prior (e.g., standard normal distribution), which is called the forward process. For the reverse process, a trained neural network is used to generate clean speech from the aforementioned prior distribution, such as SGMSE+ \cite{richter2023speech}. Besides, the appropriate condition guided generative process can achieve more superior and stable enhancement performance \cite{lemercier2023storm,yang2024pre}.

\sloppy
Although the above approaches (discriminative and generative) have achieved significantly superior performance for speech enhancement, there are still some challenges: 
\begin{itemize}
\item For discriminative methods, a deterministic map is learned during the training process. However, training data is a finite set and cannot cover all possible noise conditions to guarantee the generalizability capacity in unseen situations. Various noise types and levels can also result in distortions, especially for complex data distributions.
\item Diffusion models are always trained to learn a standard normal distribution, which requires hundreds of inference steps leading to a heavy computational cost. Additionally, the defined standard white Gaussian is also not the case of environmental noise. Although Lemercier et al. \cite{lemercier2023storm} utilizes a predictive model as a guidance, the predictive results that deviate from target distributions to a certain extent are used to compute score function, resulting in a suboptimal performance.
\end{itemize}

\sloppy
To address the aforementioned problems, we propose a multi-view discriminative enhanced diffusion-based model, named MDDM. In this work, a multi-view discriminative network is firstly used to predict an initial result that overlaps enough with a clean target distribution. Specifically in this discriminative network, the STFT-based U-Net framework is used as a backbone network. In addition, a parallel time U-Net and a noise modulation module are integrated into the backbone, which reduces speech distortions and improves noise perception abilities. To further improve the performance, we introduce a diffusion model using the discriminative intermediate feature with multi-view information as a conditioner. Furthermore, since there is an intersection distribution between target and discriminative result, they can achieve a nearly same noisy distribution by only several forward steps. Next, this noisy distribution can be converted to a more superior clean result by only several inference sampling steps, which accelerates inference speed. Assuming that discriminative results are reliable enough, the fewer sampling steps are required. Experiment results on two datasets show that our proposed method surpasses other baseline methods in terms of enhancement performance with the sampling step of only 30.

\section{Proposed Method}
\label{promet}

\sloppy
In this section, our proposed model will be introduced, as illustrated in Figure~\ref{frmw}a. First, the discriminative network takes multi-view features as inputs to get the initial output results. Next, the output results and intermediate features with multi-view fusion information guide the diffusion model during training and inference processes. Also, the discriminative and diffusion tasks are jointly optimized by a multi-task learning scheme to improve speech enhancement performance.

\subsection{Multi-view discriminative network}
\label{ssec:subhead}

\sloppy
As illustrated in Figure~\ref{frmw}b, the multi-view discriminative network contains three modules: STFT-based backbone network, waveform-base U-Net and noise perception module.
 
 \sloppy
\noindent{\bf STFT-based backbone.} Due to the superiority of U-Net in speech enhancement \cite{lv2021dccrn+,rouard2023hybrid}, we select it as the backbone and take the noisy spectrogram $ S_n $ as input, the real and imaginary parts are considered as two channels. Specifically, the noisy spectrogram $ S_n $ is fed sequentially into four convolution down-sampling encoders, a TF-GRU module and four convolution up-sampling decoders. Particularly, the TF-GRU module learns to model time and frequency relations respectively, which is similar to \cite{lv2021dccrn+}, except that GRU is adopted in our work. Note that a downsampling convolution layer after the TF-GRU module is used to transform feature dimension to 1 for integrating time branch. The decoder is built symmetrically with skip connections and outputs the predicted spectrogram $ S_p $.

\sloppy
\noindent{\bf Waveform-base U-Net.} Non-stationary noises are easily distinguished in the time domain. Therefore, integrating the time domain into the backbone is more robust for various noises \cite{zhang2021dbnet}. Besides, time and frequency are also complementary views for speech, which effectively reduces speech distortions. For the time branch, we adopt a parallel U-Net architecture that is similar to backbone. As illustrated in Figure~\ref{frmw}b, we take the time-domain signal $ T_n $ as input. The last convolution encoder outputs of two branches are added to a multi-head self-attention layer (MHA), and then the output is sent to a diffusion model and two decoders, respectively. These encoders in time and frequency branch are identical, i.e., each encoder contains two modules as shown in Figure~\ref{frmw}c. Note that time branch uses 1-D convolution and 2-D for the backbone network. 

\sloppy
\noindent{\bf Noise perception module.} In real-world scenarios, noise types are various so that training data cannot cover well. Besides, supervised noise classification is also not a good choice since public noise sets are finite and environmental noises are complex \cite{wang2020noise,hu2024noise}. Inspired by \cite{wang2018style}, in this work, we design an unsupervised fashion to learn frame-level noise. Specifically, the noisy magnitude spectrogram $ \left|S_n\right| $ is used as the input of the noise perception module. Next, four convolution layers and a bi-directional GRU are utilized to get frame-level representations. Then, a 8-head attention and 16 learnable noise templates are employed to get frame-level noise information. Additionally, a modulation parameter pair $ \left(\gamma,\beta\right) $ is obtained by two multi-layer perceptrons (MLP), respectively. Finally, for each frame, we can get a noise-related fusion given by:
\begin{equation}
\setlength{\abovedisplayskip}{6pt}
\setlength{\belowdisplayskip}{6pt}
N_{fusion}=E\ \odot\ \gamma\ \oplus\ \beta,
\end{equation}
where $ E\in\mathcal{R}^{F \times C} $ is the first encoder output of backbone, $ \gamma,\beta\in\mathcal{R}^{1 \times F} $, 
$ F $ is feature dimension and $ C $ is the number of channels. $ \oplus,\odot $ are the element-wise addition and multiplication along the frequency axis. We fuse the conditional noise module at first backbone encoder output as Figure~\ref{frmw}b shown.

\begin{table*}
\renewcommand{\arraystretch}{1.1}
\centering
\caption{Results of simulated (reverb and no reverb) and real-world datasets. Models sorted by the algorithm type, discriminative (D) or generative (G) are listed. SGMSE+ and StoRM use 50 reverse steps. Metrics higher are better and the best results are listed in bold.}
\vspace{-1.0mm}
\label{tab1}
\resizebox{0.98\textwidth}{!}{
\setlength{\tabcolsep}{4.0mm}{
\begin{tabular}{*{9}{c}}
  \toprule
  \multirow{2}*{{\bf Method}} & \multirow{2}*{{\bf Type}} & \multicolumn{3}{c}{{\bf Simulated reverb}} & \multicolumn{3}{c}{{\bf Simulated no reverb}} & {\bf Real-world}\\  
  \cmidrule(lr){3-5}\cmidrule(lr){6-8}\cmidrule(lr){9-9}
  & & {{\bf ESTOI}} & {{\bf SI-SDR}} & {{\bf MOS}} & {{\bf ESTOI}} & {{\bf SI-SDR}} & {{\bf MOS}} & {{\bf MOS}} \\
  \midrule
  {Mixture}& - & 0.47 & 8.23 & 2.43 $\pm$ 0.15 & 0.62 & 12.13 & 3.11 $\pm$ 0.13 & 2.73 $\pm$ 0.13 \\
  {HDemucs}& D & 0.70 & 15.86 & 3.09 $\pm$ 0.13 & 0.83 & 16.88 & 3.82 $\pm$ 0.15 & 3.64 $\pm$ 0.13 \\
  {BSRNN}& D & 0.77 & 16.49 & 3.27 $\pm$ 0.15 & 0.90 & 17.37 & 3.89 $\pm$ 0.13 & 3.68 $\pm$ 0.15 \\
  {MDM}& D & 0.83 & 17.58 & 3.40 $\pm$ 0.16 & 0.91 & 17.83 & 3.93 $\pm$ 0.14 & 3.77 $\pm$ 0.13 \\
 \midrule
  {SGMSE+}& G & 0.79 & 16.11 & 3.13 $\pm$ 0.15 & 0.85 & 17.22 & 3.69 $\pm$ 0.15 & 3.66 $\pm$ 0.16 \\
  {StoRM}& G & 0.82 & 17.25 & 3.31 $\pm$ 0.12 & 0.91 & 18.09 & 3.94 $\pm$ 0.13 & 3.84 $\pm$ 0.17 \\
  {{\bf MDDM}}& G & {\bf 0.86} & {\bf 18.13} & {\bf 3.49 $\pm$ 0.15} & {\bf 0.93} & {\bf 18.51} & {\bf 4.01 $\pm$ 0.15} & {\bf 3.93 $\pm$ 0.16}\\
  \bottomrule
\end{tabular}}}
\end{table*}

\subsection{Condition diffusion model}
\label{ssec:cdm}
\sloppy
Diffusion-based speech enhancement task can be considered as a subtask of conditional generation systems. In this system, clean target speech can be generated from the noisy speech by utilizing a conditional diffusion-based model. In our paper, we also design a unified framework by incorporating the multi-view condition into the diffusion-based generative model with the forward and reverse processes.


Generally, following the work \cite{richter2023speech,song2021score}, diffusion-based speech enhancement can define the forward stochastic diffusion processes as the general solution form to a linear SDE:
\begin{equation}
\label{eq2}
\setlength{\abovedisplayskip}{6pt}
\setlength{\belowdisplayskip}{6pt}
{\rm dx}_t=f\left(x_t,y\right)dt+g\left(t\right)dw,
\end{equation}
where $ x_t $ is the current state, $ y $ is the noisy condition signal, $ t\in\left[0,T\right] $ is a continuous variable describing the current $ t $-step in this process. $ w $ denotes a standard Wiener process. $ f\left(x_t,y\right) $ is called the drift coefficient, and $ g\left(t\right) $ is the diffusion coefficient, which controls the scale of the Gaussian noise injected at the current time-step $ t $. 

Furthermore, for the reverse diffusion process, it has an associated solution of the reverse SDE according to Equation \ref{eq2}, which can be defined as follows:
\begin{equation}
\label{eq3}
\setlength{\abovedisplayskip}{6pt}
\setlength{\belowdisplayskip}{6pt}
dx_t=\left[-f\left(x_t,y\right)+{g\left(t\right)}^2\nabla_{x_t}logp_t\left(x_t|y\right)\right]dt+g\left(t\right)d\bar{w},
\end{equation}
where $ d\bar{w} $ is a Brownian motion and $ \nabla_{x_t}logp_t\left(x_t|y\right) $ is the gradient term of conditional probability density distribution that can be estimated by a neural network called score model $ s_\theta\left(x_t,y,t\right) $. Finally, in inference, we can obtain the reverse SDE updated as:
\begin{equation}
\label{eq4}
\setlength{\abovedisplayskip}{6pt}
\setlength{\belowdisplayskip}{6pt}
dx_t=\left[-f\left(x_t,y\right)+{g\left(t\right)}^2s_\theta\left(x_t,y,t\right)\right]dt+g\left(t\right)d\bar{w}.
\end{equation}

Following the aforementioned procedure, in our paper, some modifies are made. Firstly, we utilize the lightweight NCSN++M architecture \cite{lemercier2023storm} as the score network but use cross attention for multi-view condition fusion and channel concatenation for predicted spectrogram $ S_p $ fusion. In addition, to train the score model in the frequency domain, at an arbitrary time step $ t\in\left[0,T\right] $, we sample to obtain the noisy spectrogram $ S_t $ from a Gaussian distribution, which can be written as follows:
\begin{equation}
\setlength{\abovedisplayskip}{6pt}
\setlength{\belowdisplayskip}{6pt}
S_t=\mu\left(S_c,S_p,t\right)+\sigma\left(t\right)z,
\end{equation}
where $ z $ is sampled from $\mathcal{N}\left(z;0,I\right) $, $ \mu $ and $ \sigma\left(t\right) $ have the identical formula as \cite{lemercier2023storm}. $ S_p $ is the predicted spectrogram of discriminative network and $ S_c $ is the clean spectrogram. Furthermore, the training objective can be written as:
\begin{equation}
\setlength{\abovedisplayskip}{6pt}
\setlength{\belowdisplayskip}{6pt}
{\rm argmin}_\theta\mathbb{E}_{t,S_p,C_m,z,S_t|\left(S_c,S_p\right)}||s_\theta\left(\left[S_t,S_p\right],C_m,t\right)+\frac{z}{\sigma\left(t\right)}||_2^2,
\end{equation}
where $ C_m $ is the multi-view condition which comes from discriminative network. The conditions of $ C_m $ and $ \left[S_t, S_p\right] $ are entered into the score model for training. Finally, after the score model is trained, we can obtain the reverse SDE according to Equation \ref{eq4} for inference.

\subsection{Training and inference}
\label{ssec:ti}
\sloppy
For the training process, we first train the multi-view discriminative network 200k steps with L1 and L2 losses. Then, we utilize the multi-task learning scheme to jointly optimize the discriminative and diffusion network until convergence. For inference, we first predict the discriminative result, and then generate the noisy sample $ S_k $ at step $ k $ through the diffusion forward process as follows:
\begin{equation}
\setlength{\abovedisplayskip}{6pt}
\setlength{\belowdisplayskip}{6pt}
S_k=S_p+\sigma\left(k\right)z.
\end{equation}
Note that $ S_k $ is not a standard normal distribution, it has the identical distribution compared with $ S_t $ in Section \ref{ssec:cdm}. Therefore, combining the intermediate multi-view feature and from $ S_k $ , the reverse diffusion process is only performed $ k $ iterations denoising to generate the clean spectrogram, and then the inversion STFT is used to get the final waveform.

\section{Experiments}
\label{exp}

\subsection{Datasets}
\label{ssec:data}

\sloppy
\noindent{\bf Training.} We use a clean 585-hour mixture dataset for model training. Specifically, 500-hour dataset is created by clean vocal track of television drama from our intranet sites. Additional 85-hour clean dataset comes from AISHELL-3 dataset\cite{shi2020aishell}. DEMAND \cite{thiemann2013diverse} and QUT-NOISE datasets \cite{dean2010qut} are selected as noise data and are split as training and testing. In addition, 10000 room impulse responses (RIRs) with T60 between 0.1 and 1.0 seconds are randomly simulated using gpuRIR method \cite{diaz2021gpurir}. In training, we randomly select clean data to convolve RIRs, and then mixed noisy data is obtained by mixing noise and reverb speech at a random uniform distribution signal-noise ratio (SNR) between 0 and 20 dB. Finally, the data distributions of noise-only, reverb-only and both are 40\%, 30\%, 30\% in mixed noisy data, respectively.

\sloppy
\noindent{\bf Testing.} In testing, we use two datasets, a simulated dataset and a real-world dataset. Specifically, for the simulated dataset, we randomly select 100 utterances from AISHELL-3 dataset for test data simulation and these utterances cannot participate in training. First, we randomly select 40\% utterances to convolve extra-simulated RIRs, and then add testing noises to all utterances with a SNR uniformly sampled from $\left[0,20\right] $ dB, which formulates the reverberation or no-reverberation test datasets. For the real-world dataset, we randomly select 20 noisy utterances of undivided track from our intranet sites, including film, drama and variety show.

\subsection{Training setups and baselines}
\label{ssec:set}

\sloppy
For discriminative network, (kernel 4, stride 2) along the frequency and (kernel 8, stride 4) along the time are used for all encoders in frequency and time network branches, respectively. The dilation factors are 1 and 2 in each encoder. For the backbone network, the time axis is no downsampling except for the last encoder (stride is 2). The first output channel is 32 and factor is 2 for both branches. The hidden units of TF-GRU are 256 and the convolution after it uses kernel 16 and stride 1 along the frequency axis. For noise perception module, the bi-directional GRU units are 128 and the dimension of noise templates is set as 256. For the diffusion, the same configuration follows \cite{lemercier2023storm}. The hop size and FFT length are 128 and 512, and the window length is 512. All training samples are resampled at 24k Hz.  We use a max of 100 epochs for all training.

\sloppy
Two discriminative methods (HDemucs \footnote{\url{https://github.com/facebookresearch/demucs}} \cite{defossez2021hybrid} and BSRNN \footnote{\url{https://github.com/sungwon23/BSRNN}} \cite{yu2022high}) and two generative methods (SGMSE+ \footnote{\url{https://github.com/sp-uhh/sgmse}} \cite{richter2023speech} and StoRM \footnote{\url{https://github.com/sp-uhh/storm}} \cite{lemercier2023storm}) are used as baselines. We train all baselines using public available codes. Besides, we also compare our multi-view discriminative model (MDM) with baselines. For MDDM, the sampling step is set as 30. For evaluation, scale-invariant signal-to-distortion ratio (SI-SDR) \cite{le2019sdr} and extended short-time objective intelligibility (ESTOI) \cite{jensen2016algorithm} are used as objective metrics and mean opinion score (MOS) \cite{streijl2016mean} is subjective metric. For MOS test, 20 samples are randomly selected from each test dataset. A total of ten people participate and participants are required to evaluate each utterance once.

\begin{table}
\renewcommand{\arraystretch}{1.1}
\centering
\caption{The results of multi-view ablation experiments on the simulated "no reverb" and real-world datasets.}
\vspace{-1.0mm}
\label{tab2}
\resizebox{0.47\textwidth}{!}{
\setlength{\tabcolsep}{2.5mm}{
\begin{tabular}{*{5}{c}}
  \toprule
  \multirow{2}*{{\bf Method}} & \multicolumn{3}{c}{{\bf Simulated no reverb}} & {\bf Real-world}\\  
   \cmidrule(lr){2-4}\cmidrule(lr){5-5}
  & {{\bf ESTOI}} & {{\bf SI-SDR}} & {{\bf MOS}} & {{\bf MOS}} \\
  \midrule
  {\bf MDDM}& {\bf 0.93} & {\bf 18.51} & {\bf 4.01 $\pm$ 0.15} & {\bf 3.93 $\pm$ 0.16}\\
  {w/o. time}& 0.86 & 17.81 & 3.93 $\pm$ 0.14 & 3.84 $\pm$ 0.15 \\
  {w/o. noise}& 0.90 & 18.25 & 3.97 $\pm$ 0.16 & 3.79 $\pm$ 0.16 \\
  {w/o. both}& 0.81 & 17.52 & 3.88 $\pm$ 0.15 & 3.74 $\pm$ 0.16 \\
  \bottomrule
\end{tabular}}}
\end{table}

\subsection{Results}
\label{ssec:result}

\sloppy
Table~\ref{tab1} shows the subjective and objective experiment results of all compared methods. Over the table, "Simulated reverb", "Simulated no reverb" and "Real-world" are the simulated test datasets with and without reverberation, as well as created real unlabeled noisy samples, respectively. "Mixture" refers to the original noisy samples. Notably, generative models generally achieve better performance in perception-related MOS compared to discriminative models, e.g., higher values of StoRM than HDemucs and BSRNN. That is because generative systems can alleviate the excessive noise suppression situation to a certain extent due to the distribution learning ability. Moreover, we extend the comparison of discriminative and generative baseline methods with our proposed method. As shown in Table~\ref{tab1}, several observations can be made. The MDDM achieves the best enhancement performance on all test datasets compared to other baselines in terms of subjective and objective metrics. Specifically, MDDM achieves the best results in terms of the ESTOI and SI-SDR on two simulated test datasets, indicating that our method can reduce speech distortions to a certain extent. In addition, MDDM obtains 3.93 MOS that is higher than other baselines on the real-world dataset, which shows powerful generalizability capacity and noise robustness. Remarkably, the multi-view discriminative model (MDM) with no diffusion process can also achieve the competitive results compared to StoRM and surpasses other discriminative methods, demonstrating the effectiveness of multi-view information fusion. Moreover, it’s worth noting that MDDM only uses the sampling step of 30 to achieve impressive performance, which greatly mitigates the computational cost compared to other diffusion-based methods using the sampling step of 50.

\begin{figure}
\centering
\includegraphics[width=\columnwidth]{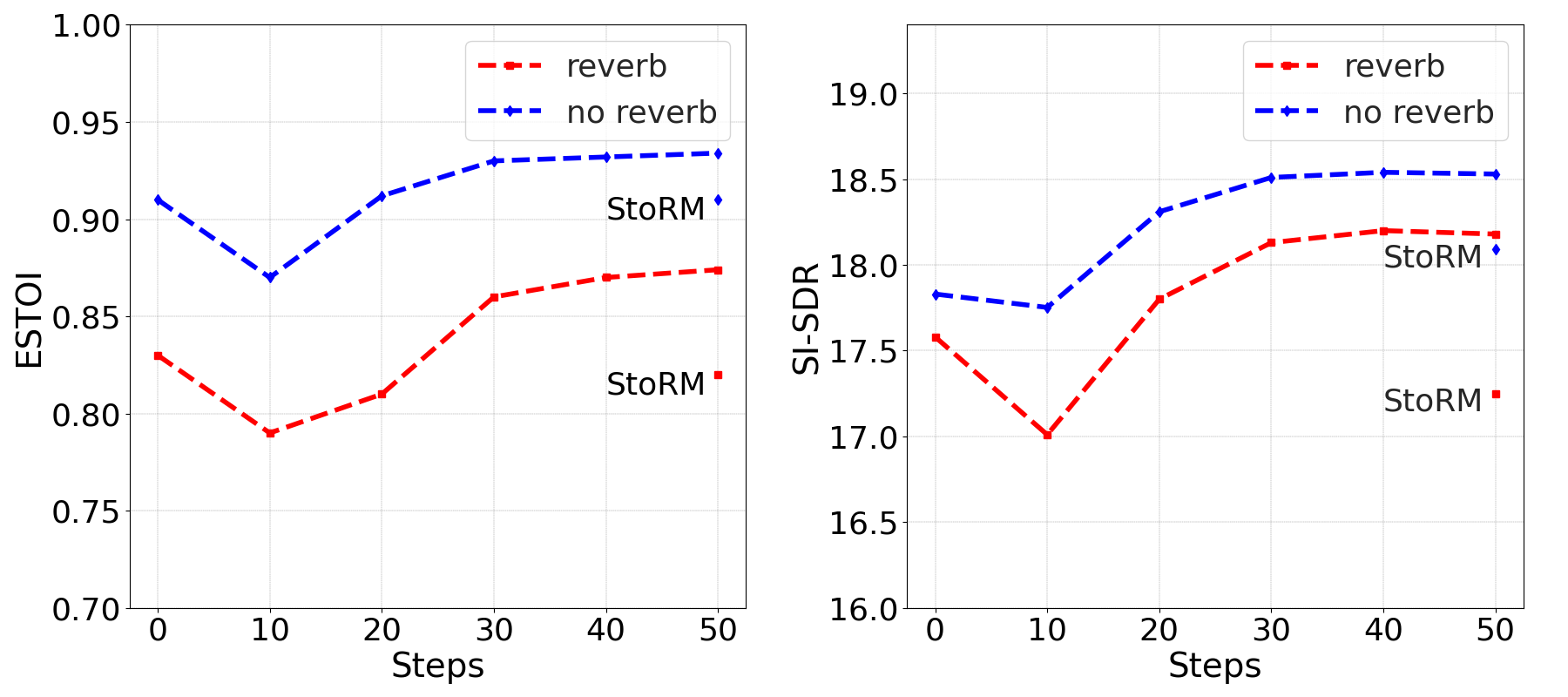}
\vspace{-0.2cm}
\caption{The objective experiment results on two simulated test datasets. The results of StoRM with the sampling step of 50 are also listed. Red and blue dots denote the simulated datasets of "with reverb" and "no reverb", respectively.}
\label{res2}
\vskip -0.1in
\end{figure}

\sloppy
As illustrated in Figure~\ref{res2}, we explore the influence of inference sampling step on the results. We utilized the objective evaluation metrics (ESTOI and SI-SDR) on two simulated datasets for testing. The max number of total step is 50 and step 0 denotes the results of MDM. Notably, as the number of the sampling step increases, the objective results also increase slowly. Additionally, as shown in Figure~\ref{res2}, although a small inference sampling step (e.g., 10) may result in the performance decline compared to MDM, the sampling step of 30 yields comparatively high quality speech, demonstrating the effectiveness of our proposed method. In contrast, StoRM \cite{lemercier2023storm} achieves the comparable performance with the sampling step of 50 and thus results in nearly 1.7 times slower than our proposed method. In our paper, for a trade-off between enhancement performance and computational burden, the inference sampling step is set as 30 for the experiments of our proposed method.

\sloppy
To verify the effectiveness of multi-view information fusion, we also perform some ablation studies in terms of conditions of the discriminative network. The time domain, noise domain and both domains are removed for testing, respectively. Table~\ref{tab2} shows the ablation results on simulated reverb and real-world datasets. Specifically, each of them removing can damage the performance. Furthermore, removing the time domain results in more speech distortions, reflected by lower ESTOI and SI-SDR values. In addition, removing the information of noise domain brings the limited generalizability capacity, reflected by a lower MOS value in real-world dataset. Therefore, the feature of each view is effective for speech enhancement so that multi-view feature can combine their strengths and thus achieve the superior enhancement performance. 

\section{Conclusion}
\label{typestyle}
\sloppy
In this paper, we propose a multi-view discriminative enhanced diffusion-based speech enhancement model. Time, frequency and noise domains are integrated into a unified framework. Multi-view feature not only generates a better discriminative output, but also can be used as a condition of diffusion-based model. Furthermore, this multi-view condition improves the generalization capabilities of enhancement model and reduces speech distortions. In addition, due to the superior enhancement performance of the discriminative network, the discriminative output has nearly identical distributions compared to the clean target. Therefore, only a small inference sampling step can be used to get final superior results, which greatly alleviates the computational burden.



\bibliographystyle{IEEEtran}
\bibliography{mybib}

\end{document}